\documentclass[final,11pt]{elsarticle}
\usepackage[a4paper, total={7in, 10in}]{geometry} 
\usepackage{graphicx} 
\usepackage{amsmath,amssymb}
\usepackage{eso-pic}
\usepackage[colorlinks=true, linkcolor=blue, citecolor=blue, urlcolor=blue]{hyperref}
\usepackage{physics}

\date{}
\begin{document}

\makeatletter
\def\ps@pprintTitle{%
   \let\@oddhead\@empty
   \let\@evenhead\@empty
   \let\@oddfoot\@empty
   \let\@evenfoot\@oddfoot}
\makeatother

\newcommand{\preprintno}{IFJPAN-IV-2025-16} 
\newcommand{\preprintdate}{September 2025} 
\AddToShipoutPictureFG*{%
  \put(\LenToUnit{\paperwidth-1cm},\LenToUnit{\paperheight-1cm}){%
    \makebox(0,0)[tr]{%
      \parbox[t]{3.2cm}{\raggedleft
        \small \preprintno\\[-2pt]
         \preprintdate
      }%
    }%
  }%
}
\title{On `$\tau$' Spin Use with \texttt{KKMCee}}

\author[ifj]{J. M. John}
\author[ifj]{Ananya Tapadar}
\author[ifj]{Zbigniew Was}

\address[ifj]{Institute of Nuclear Physics, Polish Academy of Sciences,\\
PL-31342 Kraków, Poland}

\begin{abstract}
Spin of $\tau$ represents an interesting phenomenology data point in both its production from $e^+ e^-$ collision and its subsequent decay. In precision applications such as Belle-II or FCC, where precision tagging is comparable or better than $0.1 \%$ Monte Carlo applications are usually necessary. \texttt{KKMCee}, the precision Monte Carlo program, is assuring such high precision in a broad energy range. But so far, spin effects were not easy to grasp. We introduce into \texttt{KKMCee}, \texttt{HepMC3} format output, new entries, $\tau$ helicity-like information (with weights to evaluate approximation), and $\tau$ polarimetric vectors to enable weight calculations to introduce hard interaction additional couplings. The physics details of the introduced implementations are presented, as well as examples of applications and ambiguities evaluation. The most recent updates and the useful codes of our work is provided at \href{https://th.ifj.edu.pl/kkmc-demos/}{https://th.ifj.edu.pl/kkmc-demos/index.html}. 
\end{abstract}
\maketitle
        
\section{Introduction}
The Monte Carlo simulation of the combined production and decay process of `$\tau$' has been used to look for Standard Model (SM) and beyond SM (BSM) signatures for a long time. In \texttt{KKMCee}, both spin effects and radiative corrections (including hard bremsstrahlung) are implemented. 
The analysis of spin weight in spin basis was used in the past~\cite{ALEPH:2001uca} for the measurement of the Weinberg angle `$\theta_W$' using $\tau$ polarization at $Z$ resonance. For that purpose, we need helicity-like information stored by \texttt{KKMCee}; event weights corresponding to the complete spin treatment and to helicity-like approximation.  The $\tau$ polarization near the $Z$ resonance (as the effect of ISR and FSR can be neglected ) is  approximated, as~\cite{Eberhard:1989ve},
\begin{align}
\label{eq:tau_polarization_exp}
P_\tau(\cos\theta) = -\frac{A_\tau (1 + \cos^2\theta) + 2 A_e \cos\theta}{(1 + \cos^2\theta) + 2 A_e A_\tau \cos\theta}
\end{align}
where, `$\theta$' is the angle between the directions of the incoming beam and the outgoing $\tau^-$ in the lab frame. The fitting of the parameters $A_\tau, A_e$ with experimental data provides the vector and axial vector coupling of the $Z$-boson with $\tau$. Spin weight calculation of \texttt{KKMCee} event in helicity basis has been applied to the determination of $\theta_W$ for many years; for example, Refs.~\cite{L3:1998oan,DELPHI:1999yne,CMS:2023mgq}.

Another application of precise spin weight calculation of $\tau$ production and decay is needed in search of new physics. Reweighting of spin and cross-section weights in the presence of new physics is a method particularly relevant for studying observables such as the anomalous magnetic moment or scenarios with nontrivial parity structures~\cite{Banerjee:2022sgf}. Unlike the standard precision applications, this case requires the full spin information but the precision to the added effect does not need to be very high. To enable this, \texttt{KKMCee} provide the following ingredients: the momenta of the $\tau$ leptons, their decay products, and from now on the corresponding polarimetric vectors. The matrix elements must be supplied by the user, with freedom to adopt their preferred conventions for the Standard Model and possible new-physics extensions as well as quantization frame orientation. Approximations for Bremsstrahlung may still be applied if necessary. For the kinematics, simplified treatments can be adopted—for instance, the Mustraal frame~\cite{Richter-Was:2016mal} offers a practical choice. This strategy assumes that potential new-physics interactions appear at energy scales higher than those governing Bremsstrahlung effects.  

In this article, we have provided some examples of spin weight calculation and how it affects the observables, but mostly as prototypes for experimental use, and also to explain how ambiguities of the resulting methods can be evaluated. In Sec.~\ref{sec:basics}, we outline the basic aspects of spin–weight calculation and its use for the Standard Model observable $\theta_W$, as well as the role of reference–frame choices and their impact on spin–weight calculations in the presence of possible new-physics effects (e.g., the anomalous magnetic moment of the $\tau$). Sec.~\ref {sec:numerical_result} presents the numerical results for these two directions, and the conclusions are given in Sec.~\ref {sec:summary}. Technical details are provided in~\ref{app:tech}.
\section{Basics}
\label{sec:basics}
In this section, we discuss the separation of production and decay of the $\tau$ leptons and the choice of reference frames in accordance with the observable. The production process of $\tau$ in this report is restricted to 
$e^+ e^- \to \tau^+ \tau^- (+ n\gamma)$. With an appropriate frame choice, the hard interaction can be separated into the Born part and radiative corrections as discussed in Ref.~\cite{Jadach:1998jb,Jadach:1999vf, Davidson:2010rw}. 
In this approach, the production of the lepton pair is fully encoded in the spin density matrix $R_{ij}$, while the decay kinematics enter only through the polarimetric vectors $h^\pm$. 
For ultra-relativistic $\tau$ leptons, and in the absence of sizeable quantum-entanglement effects, one may neglect terms suppressed by $m_\tau/E_\tau$ as well as the transverse components of the $\tau$ decay products’ spin analyzers. This ``helicity-only'' treatment simplifies the structure of the spin density matrix and is often adequate for SM studies at high energies.

The choice of spin quantization frame has to be adapted to the analysis goals. The so-called Mustraal frame, for instance, can be used to reduce the impact of certain bremsstrahlung photons on the kinematics of the hard process. In this frame, the polarization states of bremsstrahlung photons do not appear in the fermionic spin–density matrix, which encodes only the $\tau^\pm$ spin degrees of freedom.

\subsection{Monte Carlo elements for helicity approximation}
\label{subsec:MC_helicity}
The spin weight for the production process of $e^+ e^- \to \tau^+ \tau^- (+n \gamma) $ and subsequent $\tau$ decays into  hadrons or leptons is,
\begin{align}
\label{eq:spin_wt}
    wt &= \sum_{i,j = t,x,y,z} R_{ij}h^i_+ h^j_- \,, 
\end{align}
 where, $R_{ij}$ is the spin density matrix of two $\tau$'s, and $h^{i,j}_{\pm}$ are the normalized polarimetric vectors from the decay of the $\tau$ in the corresponding rest frame and in this frame it is assumed that the time component of the polarimetric vector $h_t^{\pm} = 1$. The weight links the spin configuration of the initial state, transmitted through the intermediate resonance of $\tau$-pair, to the properties of its decay products.  It also provides the spin correlation of a quantum-entangled $\tau$-pair.
We can again write the spin weight as,
\begin{align}
    wt_{\text{appx}} &= \sum_{i,j,a,b = t,x,y,z} R_{ij} V_i^aV_j^b h_a^+ h_b^- \nonumber \\ 
        &=  \sum_{i,j,a,b = t,x,y,z} h_a^+ V_a^i R_{ij} V_j^b h_b^- \,,
\end{align}
where, 
\begin{align}
     V_i^a &= \delta_i^a  \nonumber  \\
&=\scriptstyle{ \left[
\begin{pmatrix}1 \\ 0 \\ 0 \\ 0\end{pmatrix}
\begin{pmatrix}1 & 0 & 0 & 0\end{pmatrix}
+
\begin{pmatrix}0 \\ 1 \\ 0 \\ 0\end{pmatrix}
\begin{pmatrix}0 & 1 & 0 & 0\end{pmatrix}
+
\begin{pmatrix}0 \\ 0 \\ 1 \\ 0\end{pmatrix}
\begin{pmatrix}0 & 0 & 1 & 0\end{pmatrix}
+
\begin{pmatrix}0 \\ 0 \\ 0 \\ 1\end{pmatrix}
\begin{pmatrix}0 & 0 & 0 & 1\end{pmatrix} \right]}\,.
\end{align}
The introduction of $V_{a,b}$ helps us to separate out the production and decay amplitude. This approach is particularly useful in practice, as it allows one to treat $\tau$ production and decay separately, even in different reference frames, and subsequently boost them into a common frame~\cite{Jadach:1998wp} where the spin configuration of the produced $\tau$-pair is consistently matched to that of its decay products.

As discussed in previous works~\cite{L3:1998oan,DELPHI:1999yne,CMS:2023mgq}, the measurement of the angle $\theta_W$ from $\tau$ polarization requires the evaluation of the positive- and negative-helicity cross sections of the $\tau$ leptons. Consequently, helicity-dependent weights are also essential for studying operators sensitive to helicity, such as `$x$' or `$\omega$' (see Sec. \ref{sec:numerical_result} for details). Since the helicity is associated with spin projection on the axis $z$ only, to neglect the transverse components of spin density matrix we consider $V_i^a = \delta_i^a(\delta_i^t + \delta_i^z)$, i.e.,
in this case, we shall not use the complete basis for $V_i^a$ rather,
\begin{align}
     V_i^a &= \delta_i^a (\delta_i^t + \delta_i^z) \nonumber  \\
V&=\scriptstyle{ \left[
\begin{pmatrix}1 \\ 0 \\ 0 \\ 0\end{pmatrix}
\begin{pmatrix}1 & 0 & 0 & 0\end{pmatrix}
+
\begin{pmatrix}0 \\ 0 \\ 0 \\ 1\end{pmatrix}
\begin{pmatrix}0 & 0 & 0 & 1\end{pmatrix} \right]} \nonumber \\
&= \frac{1}{2} \scriptstyle{ \left[
\begin{pmatrix}1 \\ 0 \\ 0 \\ 1\end{pmatrix}
\begin{pmatrix}1 & 0 & 0 & 1\end{pmatrix}
+
\begin{pmatrix}1 \\ 0 \\ 0 \\ -1\end{pmatrix}
\begin{pmatrix}1 & 0 & 0 & -1\end{pmatrix} \right]} \nonumber \\
&= \frac{1}{2}(s^+s^{+^T} + s^-s^{-^T}) = \frac{1}{2}\sum_{i=\pm} s^i s^{i^T}
\end{align} 
where $s^{\pm}$ resembles the pure spin state at very high energy with spin projection along $z$-axis.
On this basis, neglecting the transverse components we can write the approximated spin weight,
\begin{align}
     wt_{\text{appx}} &= \sum_{i,j,a,b = t,z} h^a_+ V_a^i R_{ij} V^j_b h^b_- \nonumber \\
     &= \frac{1}{4}\sum_{i,j,a,b = t,z} h^a_+ \sum_{m= \pm} (s^m)_a (s^{m^T})^i R_{ij} \sum_{n=\pm}  (s^n)_b (s^{n^T})^j h^b_- \nonumber \\
     &= \frac{1}{4}\sum_{m= \pm} \sum_{n=\pm}\sum_{i,j,a,b = t,z} \left(h^a_+  (s^m)_a \right) \left((s^{m^T})^i R_{ij}
     (s^{n^T})^j \right) \left( (s^n)_b h^b_- \right) \nonumber \\
     &= \frac{1}{4}\sum_{m= \pm} \sum_{n=\pm} (h_+ \cdot s^m) (s^m \cdot R \cdot s^n) (h_- \cdot s^n)
    \label{eq:appx_spin_wt}
\end{align}
We can see that once transverse elements of $R_{ij}$ are dropped, the picture of production of $\tau$ pair in spin (helicity) states, and their decays can be separated out. At this approximation level, we can thus attribute helicity-like quantities for individual $\tau$s. For this purpose, new attributes are added to \texttt{HepMC3} event record.  
Technical details of this construction within \texttt{KKMCee} code is given in~\ref{app:attributes_saved_hepmc3}.

\subsection{Monte Carlo elements for anomalous couplings}
\label{subsec:MC_anomalous_coupling}
In evaluating possible new couplings, the SM component must retain full numerical precision, as it forms the reference for detecting deviations. The new interaction, expected to be small or absent and requires only moderate precision, since the goal is typically to set upper bounds rather than achieve high-accuracy determination. 
We extract the polarimetric vectors of the $\tau^\pm$ decays from the \texttt{KKMCee} \texttt{HepMC} output (technical details are discussed in~\ref{app:attributes_saved_hepmc3}), while approximating the hard-process kinematics, including bremsstrahlung, and employing a simplified SM amplitude. 
This approximation reproduces \texttt{KKMCee} results well for both soft- and hard-photon scenarios at a center-of-mass energy of $10.58~\mathrm{GeV}$ as shown in Ref.~\cite{Banerjee:2022sgf}. 
Another justification of the approximation is that it is applied consistently in both numerator and denominator of the weight $\frac{\mathrm{SM+NP}}{\mathrm{SM}}$, ensuring that most ambiguities cancel.  The remaining uncertainty affects only the NP term, where the required numerical precision is modest given the expected smallness of the effect.

For the weight calculations, users need only to take from \texttt{HepMC3} event record  with the following four momenta:
electron, positron beams, $\tau^+$, $\tau^-$ momenta and polarimetric vectors $h^+$, $h^-$, respectively. All these are defined in the laboratory frame. To study the angular distribution of $\tau^\pm$ decay products, the four-vectors can be boosted to the desired reference frame. 
Here we discuss two possible variants of reference frames: 

\textbf{Collins-Soper Frame:} 
As described in Ref.~\cite{Banerjee:2022sgf}, a commonly used reference frame to avoid the complication of ISR and FSR photon kinematics in defining hard-scattering angles, the $\tau$-pair rest frame is considered. In this frame, the momenta of the outgoing lepton pairs are opposite to each other, and this is considered as the $z$-axis ( momentum of $\tau^-$ is along $+z$-axis)\footnote{Other variants of this frame are also available in the literature~\cite{Richter-Was:2016mal}, for example in the original paper \cite{Collins:1977iv} z axis is along the bisection of beams direction, and orientation of $\tau$ rest frames is not of a concern. }. To construct the other two axes, we assume that the difference in the beams' directions
is in $x-z$ plane and $\hat{y} = \hat{z} \times\vec{x}$,
finally $\hat{x} = \hat{y} \times\hat{z}$.
The net photon–recoil vector lies entirely in the $x$–$z$ plane, the $\tau^+\tau^-$ pair has no net transverse momentum in its rest frame, and the $\tau^\pm$ momenta are aligned with the $z$-axis. In this configuration, the azimuthal angle $\phi$ of the $\tau$ momentum is undefined, as its transverse projection onto the plane perpendicular to $z$ vanishes. Nevertheless, $\phi$ remains well-defined for the $\tau$ decay products, which in general carry nonzero transverse components even when the parent $\tau$ does not. In our analysis, $\phi$–dependent observables are therefore constructed from the momenta of the decay products rather than the $\tau$ momenta themselves, which we will discuss later in subsection~\ref{subsec:anomalous_numerics}.

\textbf{Mustraal frame:} \\
The properties of QED matrix elements as presented in the Ref.~\cite{Jadach:1998wp, Barberio:1993qi}. It is important to choose the reference frame in accordance.  We choose the $+z$-axis along $\tau$ direction in $\tau$ pair rest frame as in the Collins-Soper case. The rest of the verses are fixed 
are also fixed as in the Collin-Soper case.
However, now we use either the first or the second beam direction to determine the $x-z$ plane. The probabilities for the use of the first/second beam direction according to the formula,
\begin{align}
w_1 &= \frac{E_{\rho_1}^2 \left(1 + \cos^2\theta_1\right)}{E_{\rho_1}^2 \left(1 + \cos^2\theta_1\right) + E_{\rho_2}^2 \left(1 + \cos^2\theta_2\right)} \,, \nonumber \\
w_2 &= \frac{E_{\rho_2}^2 \left(1 + \cos^2\theta_2\right)}{E_{\rho_1}^2 \left(1 + \cos^2\theta_1\right) + E_{\rho_2}^2 \left(1 + \cos^2\theta_2\right)} = 1 - w_1 \,.
\label{eq:probs}
\end{align}
This fomula is as eq. 9 from \cite{Richter-Was:2016mal}. See there for details and notation. Selection of frame choice out of these two has been added now within the code (details can be found in~\ref{app:stand_alone}).

Independent of the choice of frames scheme, we calculate the $R_{ij}$ matrix for the pair of $\tau$ rest frames. The polarimetric vectors were generated by \texttt{TAUOLA} in the rest frame of $\tau$. 
It is convenient that \texttt{HepMC3} stores them after \texttt{KKMCee} boost them  to lab frame. Application working on \texttt{HepMC3} content needs to boost them back to the $\tau$ pair rest frames, only no need to worry about conventions used in \texttt{KKMCee}. Implementation of this within \texttt{KKMCee} has been discussed in detail in the~\ref{app:stand_alone}.
Now the polarimetric vectors are again in the rest frame of $\tau$ but of distinct, user-controlled, orientation.  We follow the previously explained boost and rotations so that we get $h^+_i$, $h^-_j$ and $R_{ij}$ in consistent frames: respective $\tau$ rest frames and CM frame.

\section{Numerical results}
\label{sec:numerical_result}
\subsection{Use of helicities Numerics}
\label{subsec:helicity_numerics}
\begin{figure}[ht]
    \centering
    \includegraphics[width=0.4\linewidth]{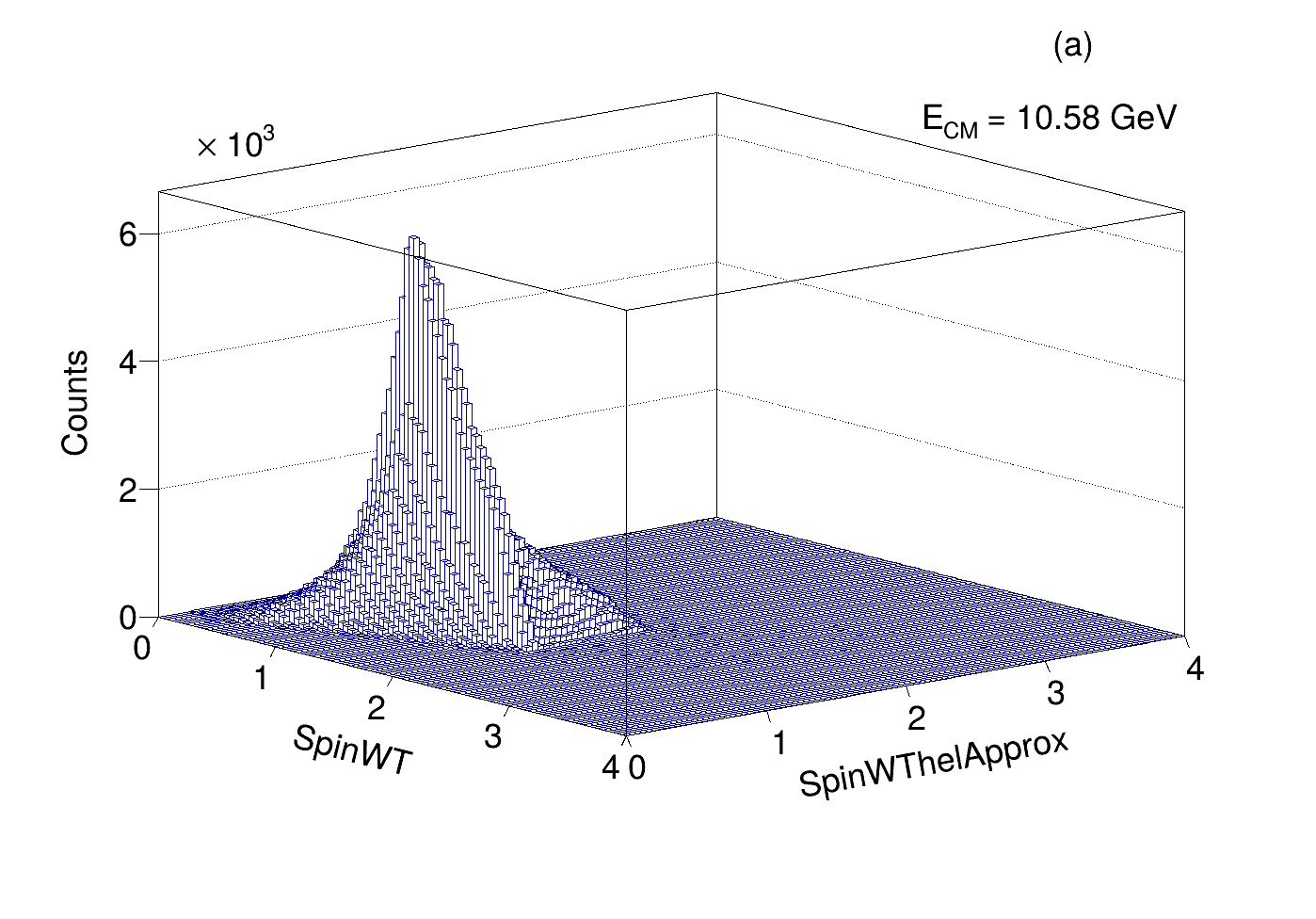}
    \includegraphics[width=0.4\linewidth]{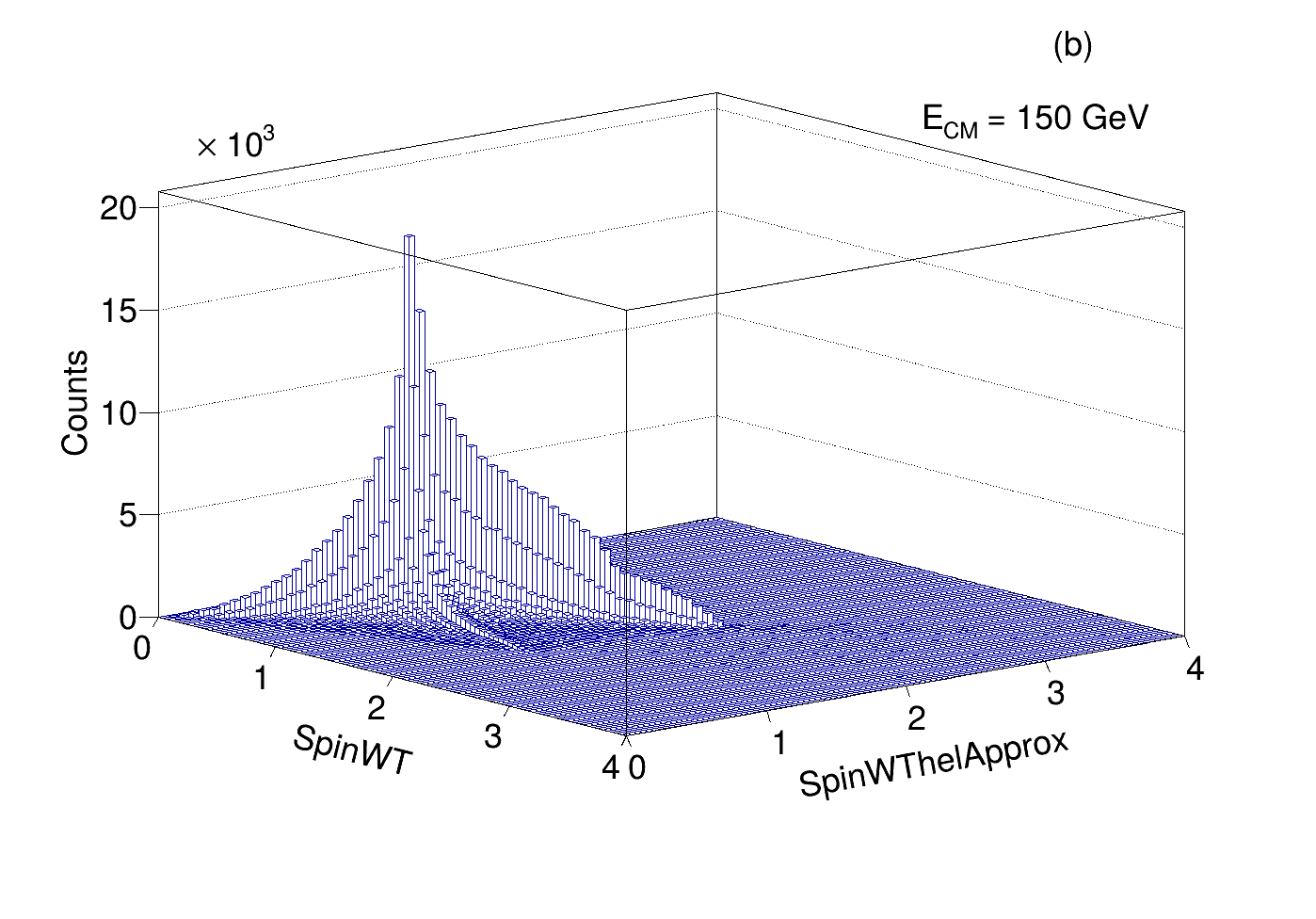}
    \caption{The correlation between complete spin weight and approximated spin weight for two different energies $10.58$ GeV (left plot) and $150$ GeV (right plot) is presented.}
    \label{fig:wt_compare}
\end{figure}

In this section, we evaluate the ambiguities of the algorithm presented in subsection~\ref{subsec:MC_helicity} by looking into the spin weight calculation. As an example, we have presented  
in Fig.~\ref{fig:wt_compare}, the correlation between two versions of the spin weight calculation (including Bremsstrahlung and virtual corrections) as given in Eq.~\ref{eq:spin_wt} and Eq.~\ref{eq:appx_spin_wt}. Here, \textit{SpinWT} denotes the weight calculated using the complete spin correlation matrix $R_{ij}$, which includes both longitudinal and transverse polarization components of the $\tau^\pm$. In contrast, \textit{SpinWThelApprox} represents an approximate weight obtained by neglecting the transverse components of $R_{ij}$, i.e., retaining only the time-like and longitudinal ($t$ and $z$) contributions.

The left panel corresponds to a center-of-mass (CM) energy of $\sqrt{s} = 10.58\ \mathrm{GeV}$, matching the Belle II operating energy, while the right panel corresponds to a higher CM energy of $\sqrt{s} = 150\ \mathrm{GeV}$. The correlation between the full and approximate spin weights demonstrates that at high energies, the absence of transverse spin correlations leads to less deviation from the full calculation than at lower energies. Even more important, at high energies, the transverse components of $\tau$ decay products momenta are with increasing energy, less and less measurable.

This behavior can be understood from the helicity structure of the $\tau$ wave functions. In the ultra-relativistic limit ($E \gg m_\tau$), the $\tau$’s spin state is well approximated by a pure helicity eigenstate, and the corrections are suppressed by the factor $\mathcal{O}(m_\tau/E)$, or its square. At lower energies,  $m_\tau/E$ is non-negligible and affects the spin weight more sizably.

Both at low and high energies, transverse components of $R_ij$ are large, but at high energies, their omission has a smaller effect on the observable distributions. Longitudinal correlation and the longitudinal polarization dominate the observable asymmetries.

Both at high and low energies, although the transverse components of $\tau$ decay products  are individually suppressed, the kinematic enhancement of spin correlations in production and decay channels means that their absence can still significantly bias precision measurements, particularly for observables sensitive to azimuthal correlations or CP-violating effects. This is a direct consequence of the fact that the transverse components in $R_{ij}$ encode spin–spin interference terms, which can survive integration over phase space and influence angular distributions.

Therefore,  the helicity approximation (SpinWThelApprox) may be adequate for studies of observables relying on particles energies. This may be good even for  precise determination of $Z$ boson polarization-sensitive quantities — such as the $\tau$’s electroweak couplings or the weak mixing angle from angular asymmetries. This does not require retaining the full spin correlation matrix, including the transverse components. However this means that subtle but essentially measurable spin–spin  effects are ignored across the full energy range. 

\begin{figure}[ht]
    \centering
    \includegraphics[width=0.29\linewidth]{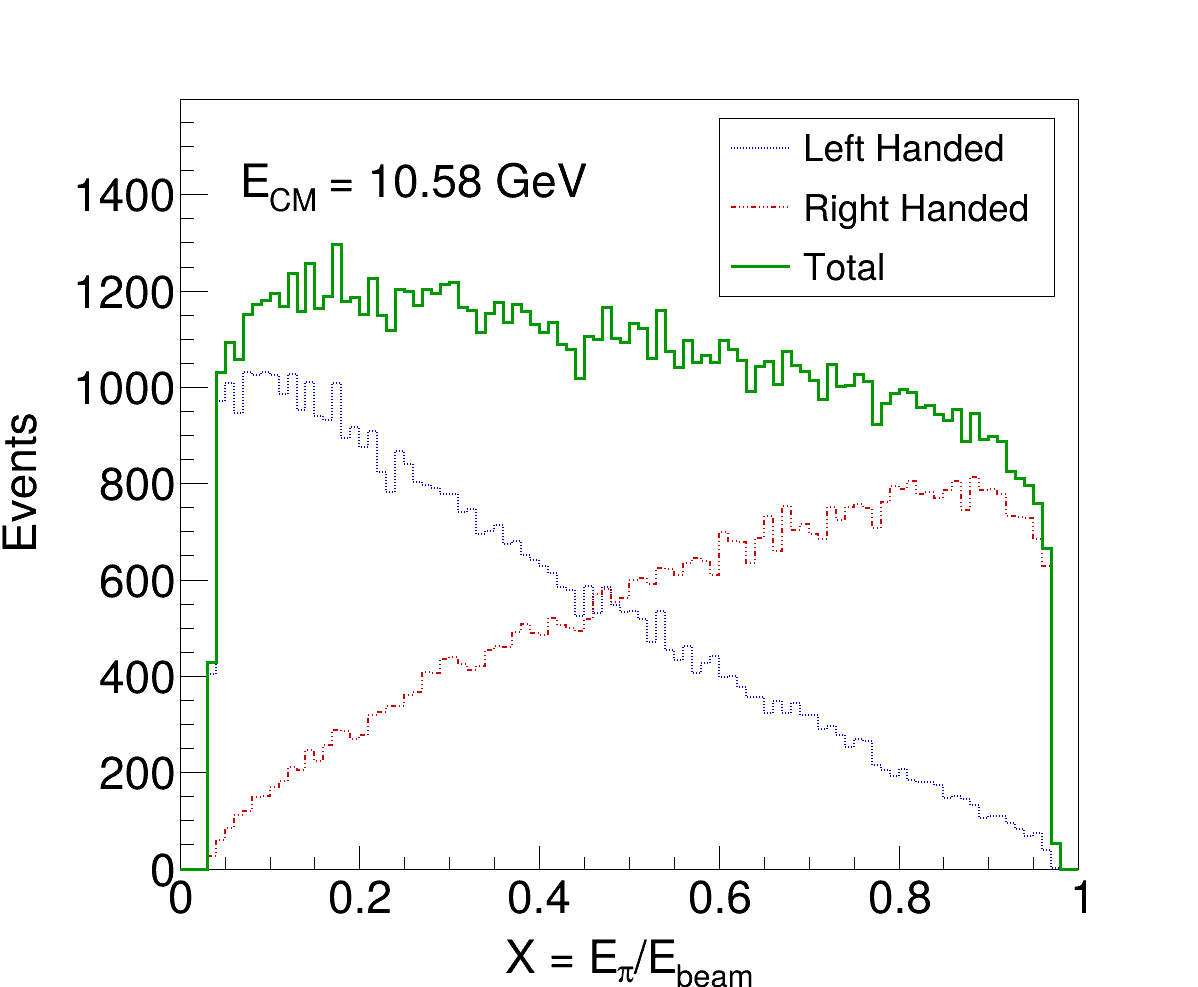}
    \includegraphics[width=0.29\linewidth]{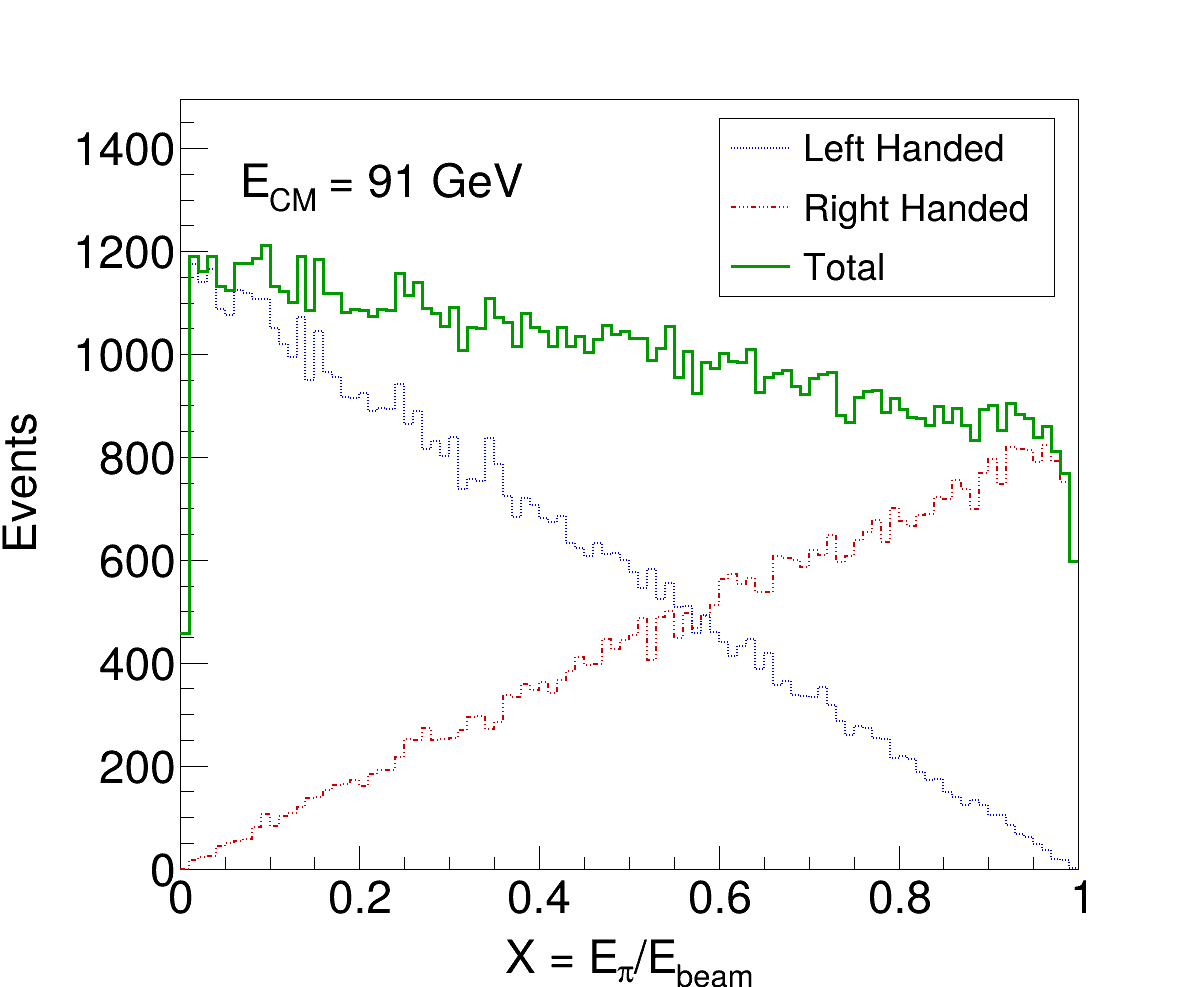}
    \includegraphics[width=0.29\linewidth]{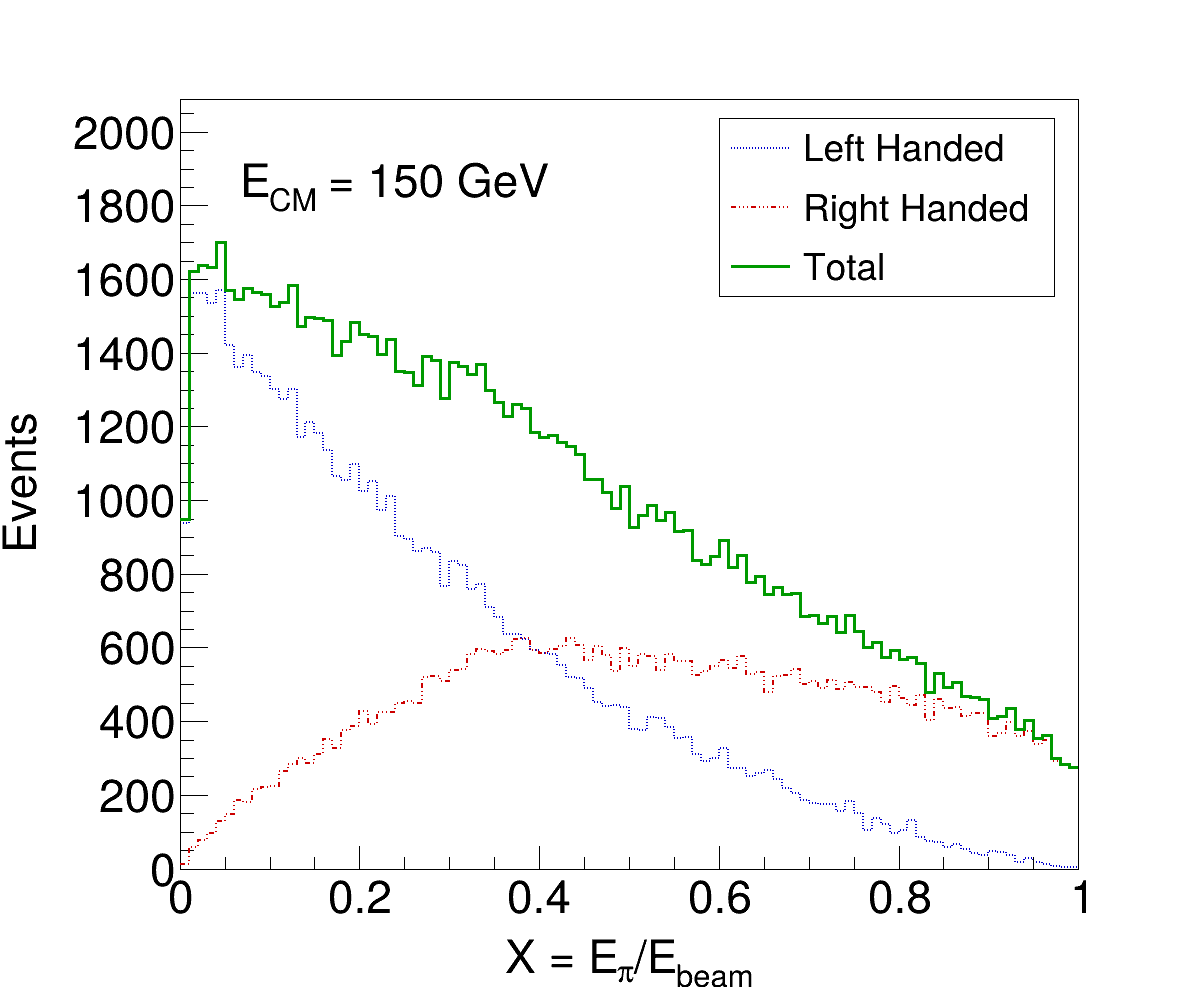}
    \caption{Pion spectrum (green line) separated into left handed (blue) and right handed (red) sub-samples. Left plot is for CM energy $10.58$ GeV, central plot $91.1842$ GeV, right plot for $150$ GeV. Initial and final state radiations are switched on that is why plots from pion spectrum deformed from simple slope shapes. First beams at $10.58$ GeV plot are empty because of pion mass. }
    \label{fig:helicity_xobs}
\end{figure}

The algorithm for the use of helicities follows the one used already in the ALEPH collaboration `$Z$'  boson couplings measurements. There is no need to repeat that work. Here we only provide the example plot~\ref{fig:helicity_xobs} for $\tau^\pm \to \pi^\pm \nu (\pi^\pm \pi^0 \nu)$ where the sample is separated into left and right components is given. This is the similar to fig. $13$ from \cite{ALEPH:2001uca}. 

In the case of $\tau^\pm \to \pi^\pm \nu_\tau, \ell \overline{\nu}_\ell \nu_{\tau}$, the charged pion momentum in the lab frame is helicity sensitive. This is because of the left-handedness of the neutrino. Therefore, in the case of single hadrons or the leptonic decay of $\tau$, helicity information is also carried by the momentum of the charged particles. Therefore, in these cases we look into the observable, $x = \frac{E_{\pi,\ell}}{E_{\text{beam}}}$. In the lab frame the energy of pions has the information of $\tau$ polarization.

 In the first panel of Fig.~\ref{fig:helicity_xobs}, we observe a reduction in the number of events when the pion energy falls below the $\pi$ mass. More generally, across all three panels, the event yield decreases with increasing $x$ for a fixed CM energy. This behavior originates from the left-handed nature of neutrinos, which biases the pion emission towards lower energies in the laboratory frame (equivalently, in the $\tau$-pair rest frame). Bremsstrahlung further reduces the effective pion energy, thereby contributing to the suppression of large `$x$' events. The handedness of the neutrino also determines the relative behavior of the distributions for positive and negative $\tau$ helicities, shown in blue (left-handed) and red (right-handed).

\begin{figure}[ht]
    \centering
    \includegraphics[width=0.29\linewidth]{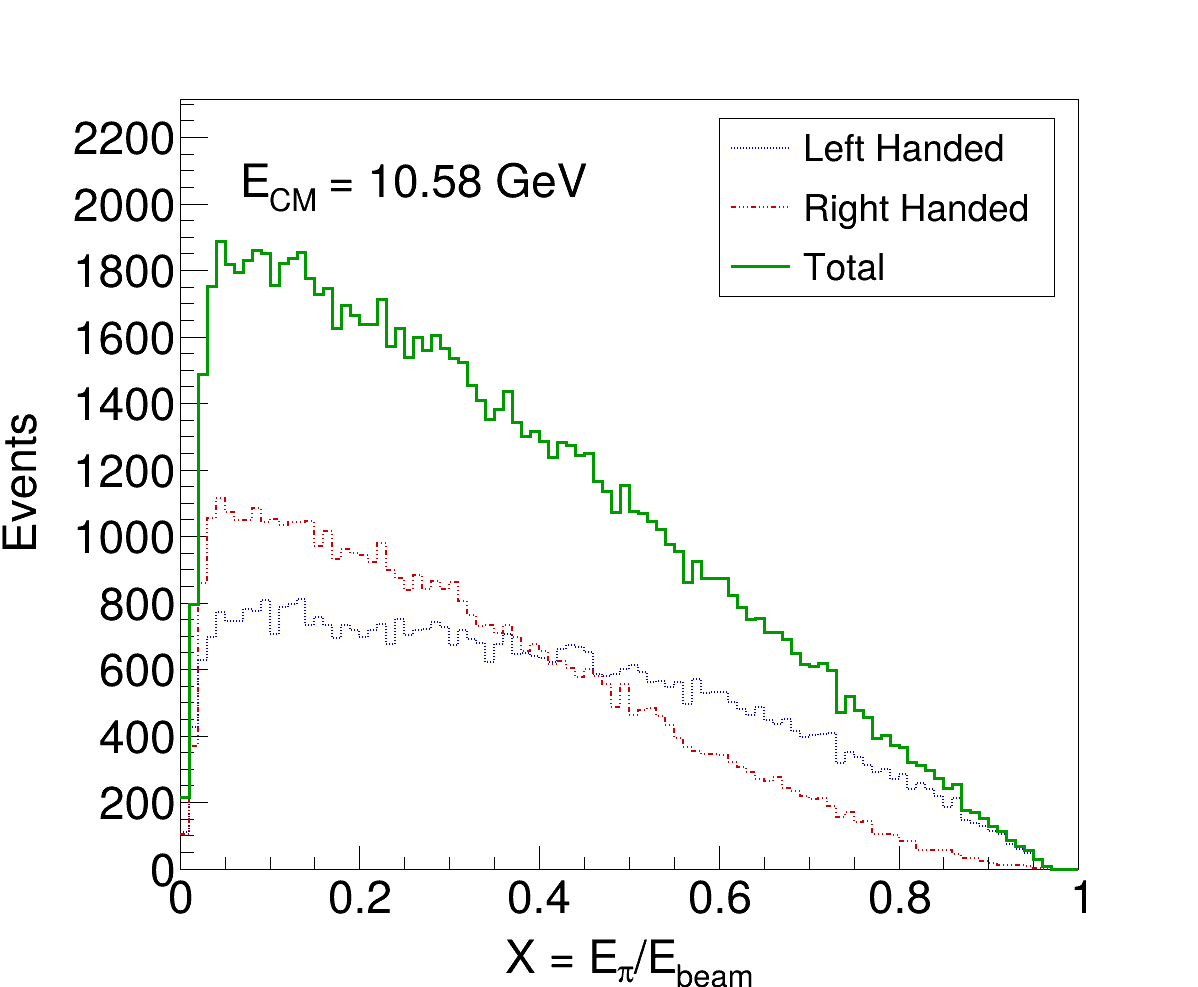}
    \includegraphics[width=0.29\linewidth]{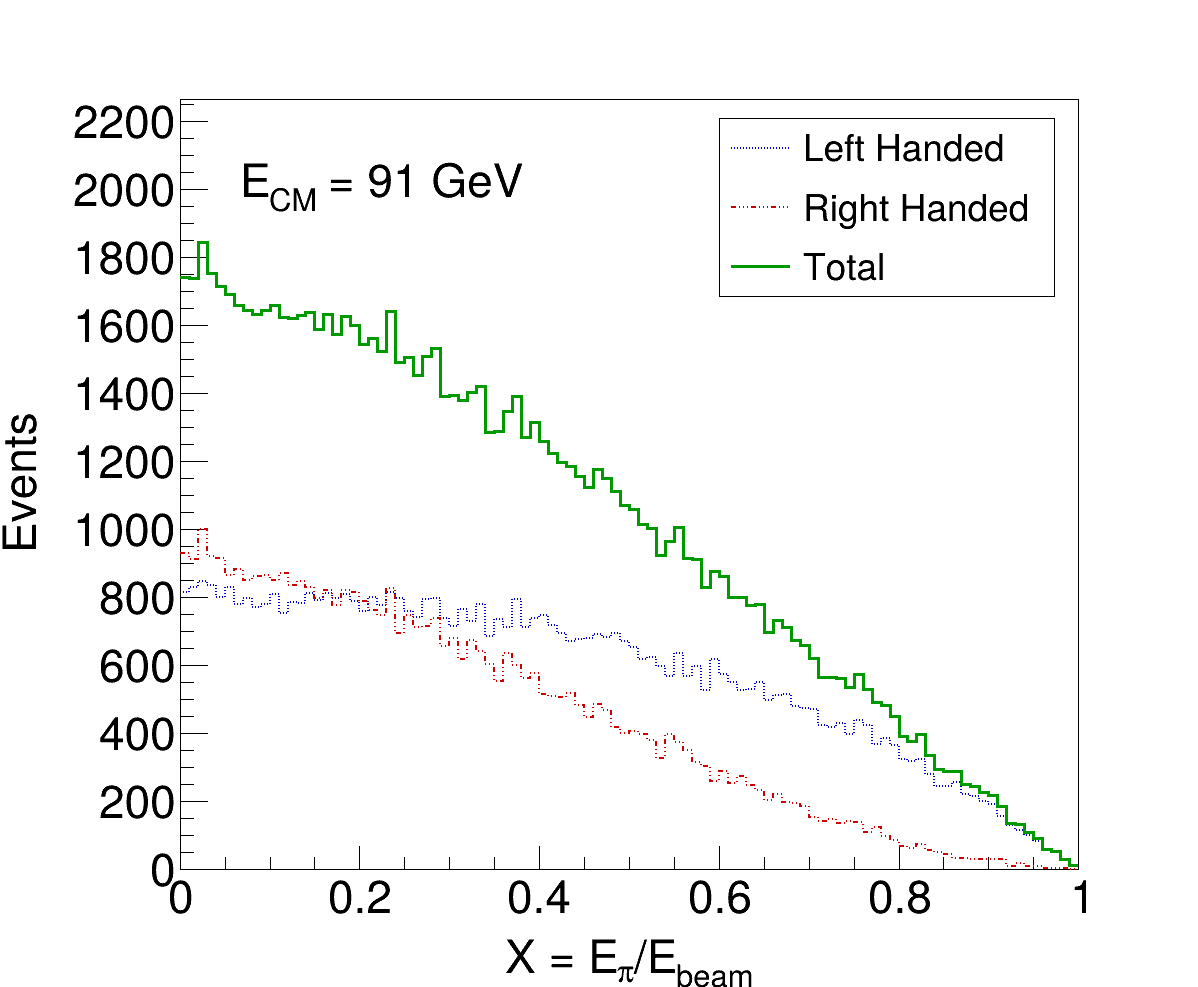}
    \includegraphics[width=0.29\linewidth]{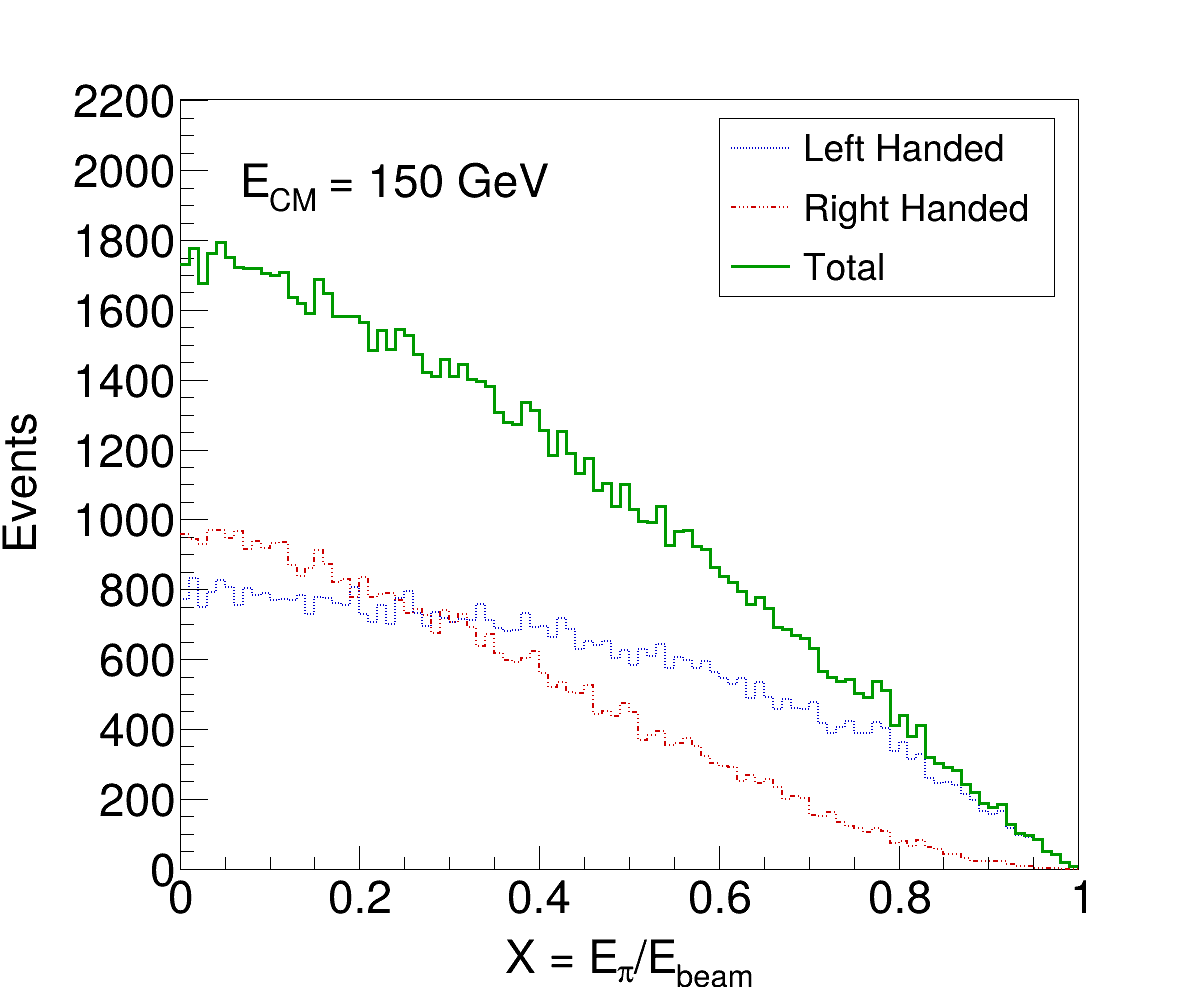}
    \caption{Electron spectrum (green line) separated into left handed (blue) and right handed (red) sub-samples. Left electron is for CM energy $10.58$ GeV, central plot $91.1842$ GeV, right plot for $150$ GeV. Initial and final state radiations are switched on that is why plots from electron spectrum deformed from simple slope shapes. First beams at $10.58$ GeV plot are empty because of electron mass. }
    \label{fig:helicity_xobs_e}
\end{figure}

\begin{figure}[ht]
    \centering
    \includegraphics[width=0.29\linewidth]{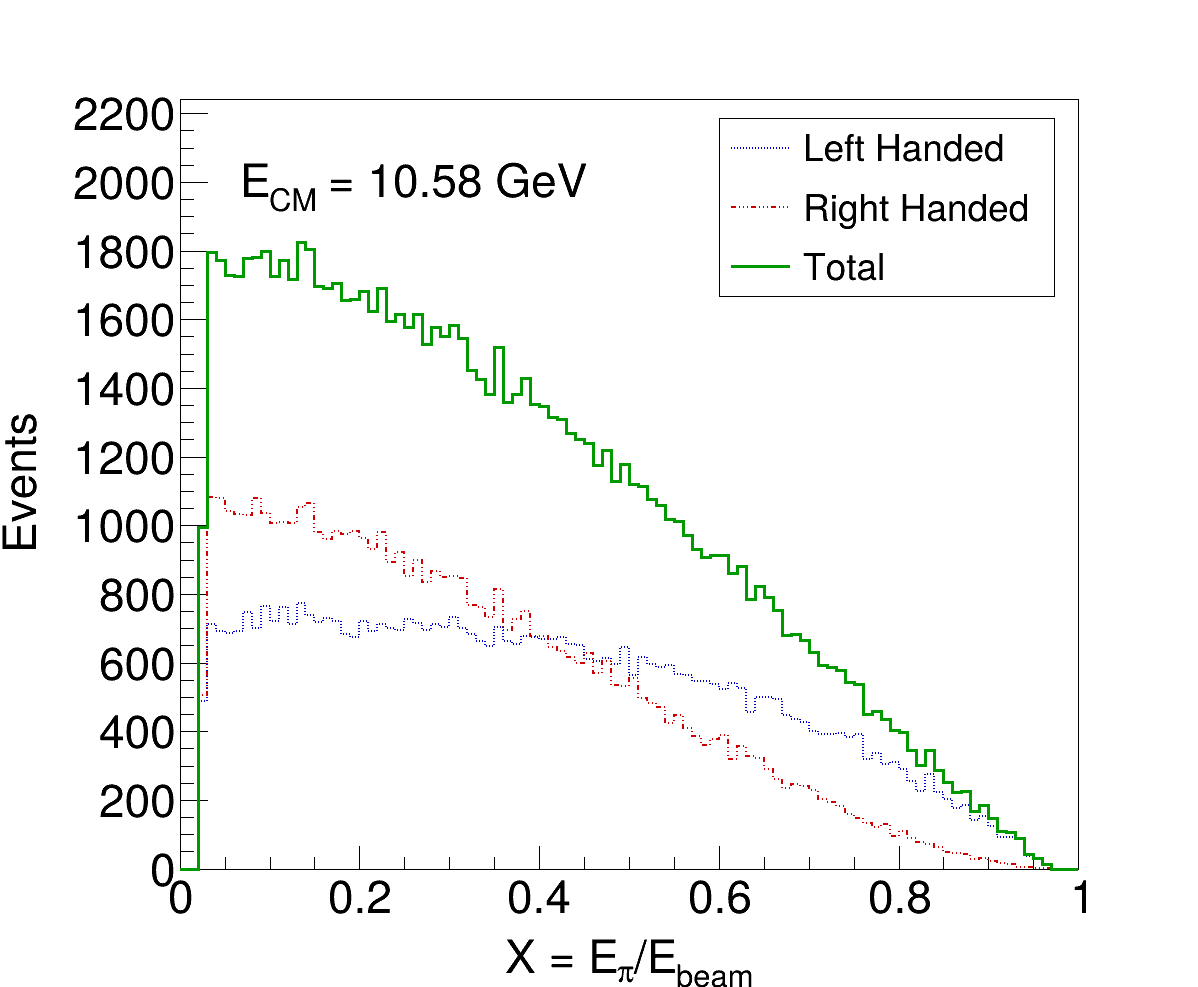}
    \includegraphics[width=0.29\linewidth]{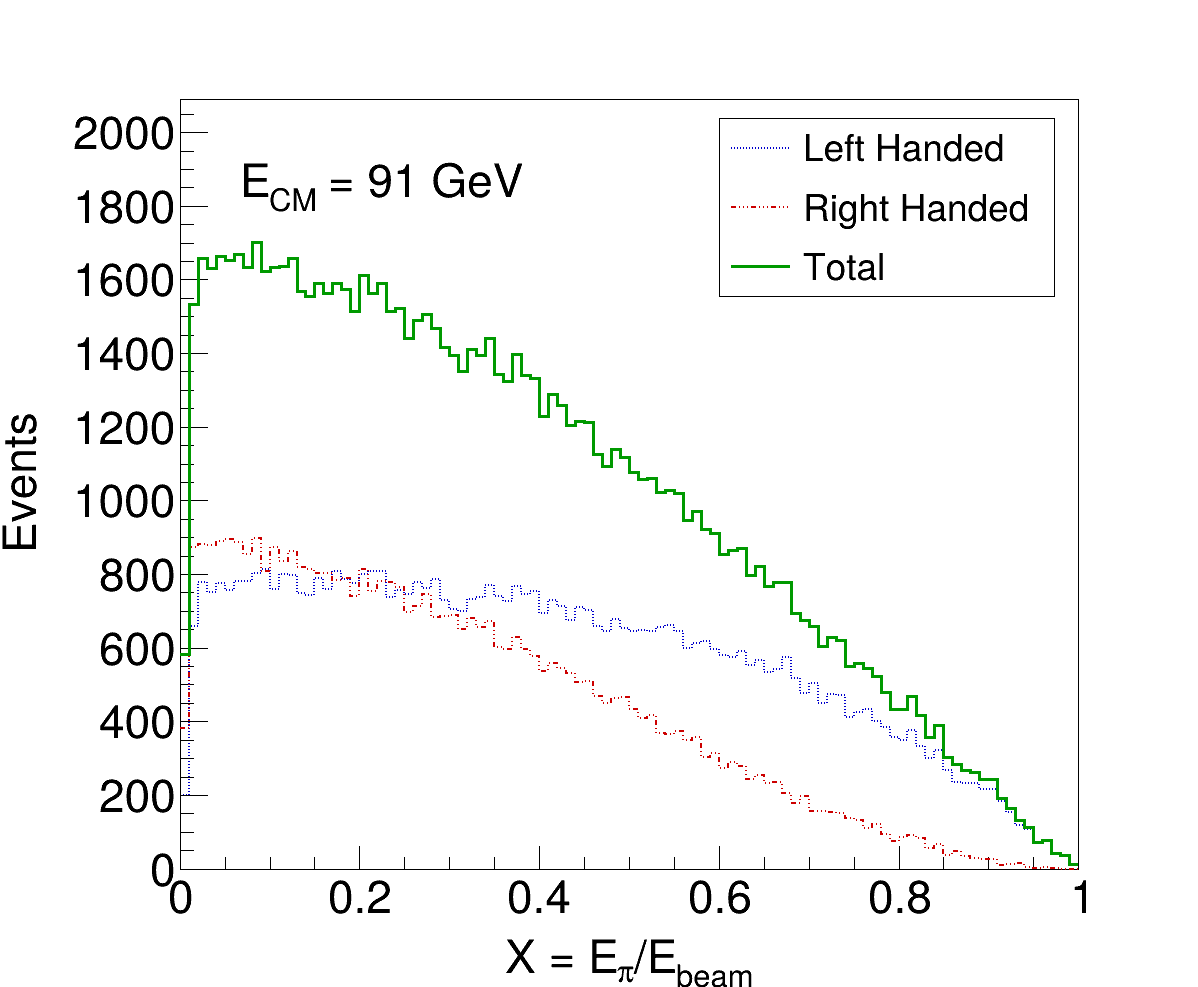}
    \includegraphics[width=0.29\linewidth]{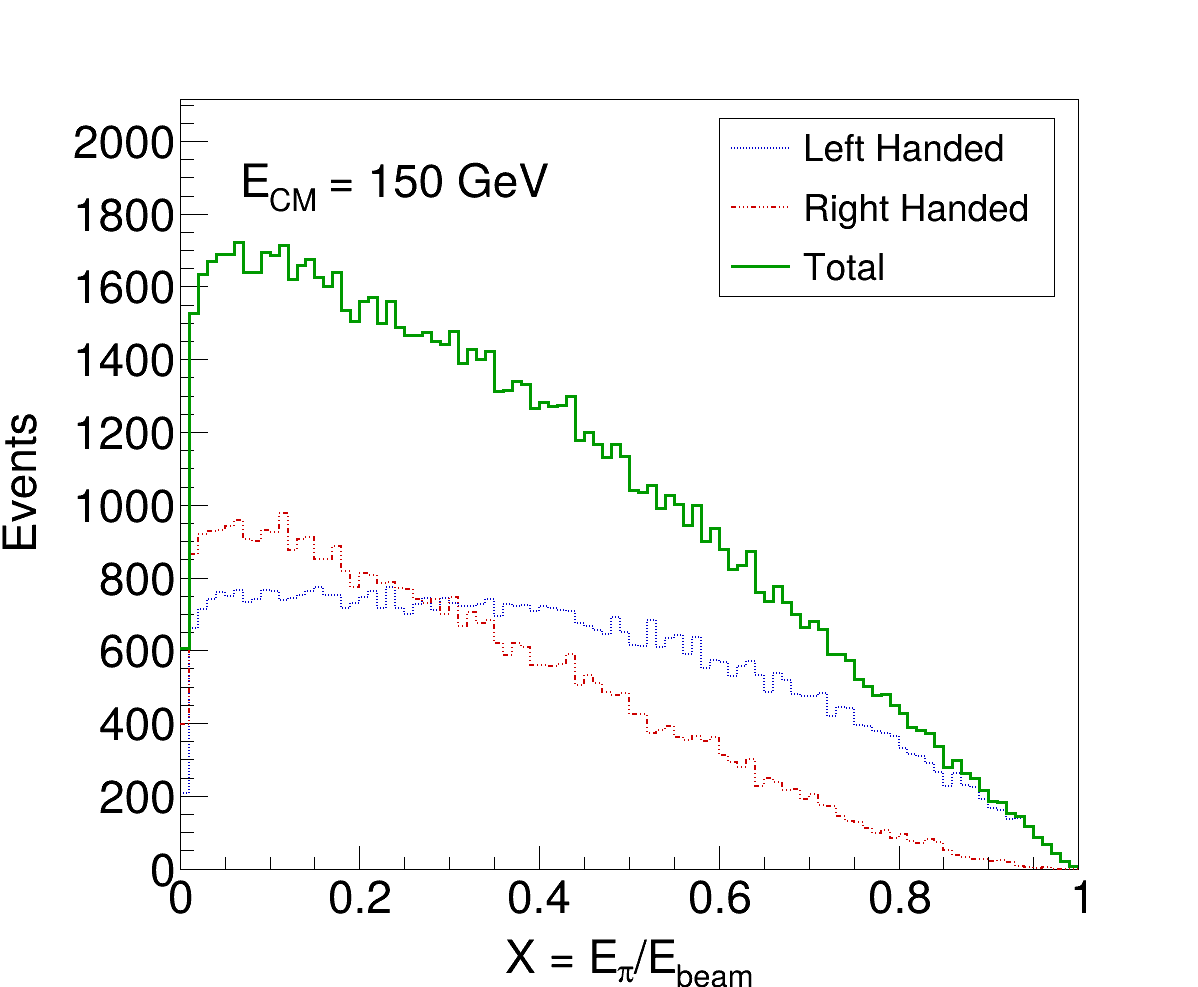}
    \caption{Muon spectrum (green line) separated into left handed (blue) and right handed (red) sub-samples. Left muon is for CM energy $10.58$ GeV, central plot $91.1842$ GeV, right plot for $150$ GeV. Initial and final state radiations are switched on that is why plots from muon spectrum deformed from simple slope shapes. First beams at $10.58$ GeV plot are empty because of muon mass. }
    \label{fig:helicity_xobs_mu}
\end{figure}

The impact of these effects can be understood by comparing the first two panels. In the second panel, the CM energy is set to $91$ GeV, close to the $Z$-boson mass, where the cross section is dominated by the $Z$ resonance. As a result, ISR and FSR contributions have little influence on the total number of events. By contrast, at higher CM energies, ISR corrections become increasingly important: as expected from QED, they modify the Born cross section logarithmically, $\sim \frac{\alpha_e}{\pi} \ln(s/m_e^2)$. Consequently, the number of events at large pion energies is significantly reduced, leading to the sharp fall observed in the third panel.
We have also regenerated the events for leptonic decay of $\tau^\pm$ in fig.~\ref{fig:helicity_xobs_e}(for $\tau^- \to e^- \bar{\nu}_e \nu_\tau$), fig.~\ref{fig:helicity_xobs_mu} (for $\tau^- \to \mu^- \bar{\nu}_\mu \nu_\tau$).

However, for more than two-body decay of $\tau$  into hadrons, we have to consider another variable, for example  $\omega$ as defined in Ref.\cite{ALEPH:2001uca}

With the increasing precision of future measurements, tests of the application limit using ratio of spin weigth and spin weight truncated to helicity level should be sometimes considered\footnote{Note that we have to take the ratio, because usually in our sample the spin effects are taken in. That is the way to check for ambiguties of the approximation but technically ratio of the weight is not bound from above, there may be integrable infinities. If this is the problem, events sample should be generated without spin, then spin weights would be used respectively for full spin and helicity truncated spin. Both weights would be then in the range $0$ to $4$ and of average $1$.} nonetheless for some observables such as `$x$' transverse components of $\tau$ polarization up to certain precision level  can be ignored.
Such ambiguity evaluation requires detection details and will also be $\tau$ decay channel dependent.
But the plots of our fig \ref{fig:helicity_xobs} are enough to validate our re-coding of the solution developed a long time ago in \texttt{F77}.

\subsection{Use of polarimetric vector}
\label{subsec:anomalous_numerics}
\begin{figure}[ht]
    \centering
    \includegraphics[width=0.9\linewidth]{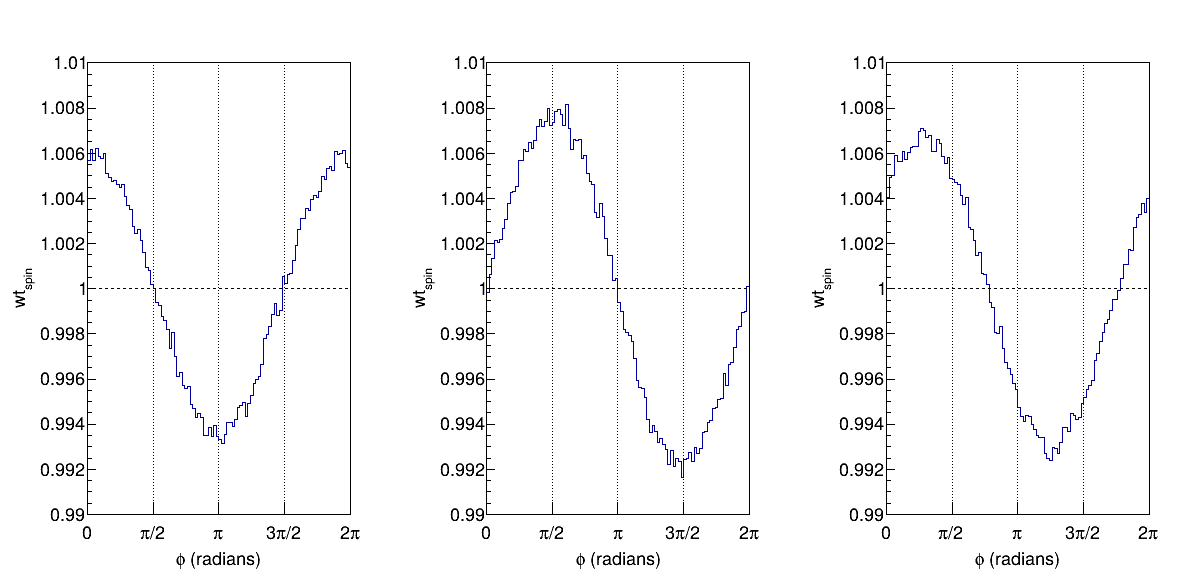}
    \caption{In this figure, which reproduce the fig.2 of Ref.~\cite{Banerjee:2022sgf} (see this reference for details) we present the variation ratio of the spin weight with and without new physics coupling for the acoplanarity angle between $\tau$ decay products. In the left plot, $\mathrm{Re}(a_{\text{NP}})=0.04$, in the middle plot, $\mathrm{Re}(b_{\text{NP}})=0.04$ and in the right most plot, $\mathrm{Re}(a_{\text{NP}})=0.04\cos(\pi/4)$ and $\mathrm{Re}(b_{\text{NP}})=0.04\sin(\pi/4)$, with the remaining anomalous couplings set to zero. }
    \label{fig:anomalous_moment}
\end{figure}

Let us follow another example, this time for the use of polarimetric vectors $h^\pm$. 
For that purpose, we will reconstruct some plots of ref.\cite{Banerjee:2022sgf} which were prepared in \texttt{FORTRAN}.
The algorithm of reweighing is essentially the same as explained in that paper.  The use of the anomalous dipole moments paper can be treated as an example. The difference is that previously, the polarimetric vectors were presented in the $\tau$ rest frame, and the user has to provide proper rotation of frames to get the polarimetric vector in the required frame. In this new framework, we have a polarimetric vector in the lab frame in \texttt{HePMC3} output. As a consequence, the user can boost the polarimetric vectors to any reference frame needed according to the observable without worrying about the rotation of frames within \texttt{KKMCee}. A new standalone code in \texttt{C++} is developed where the event inputs are obtained from \texttt{HepMC3} output file (with the information of spin weight, polarimetric vectors, decay products) in the lab frame.
We have checked that the weight factor event by event from these two codes matches up to $0.1 \%$ when Bremsstrahlung processes are not present (or are soft). That is the first test of the correctness.

To validate translation from \texttt{f77} to \texttt{C++} of the full application, we redo some distributions, see Fig.~\ref{fig:anomalous_moment} now with the $C++$ application.
We consider the same decay channel, $\tau^- (\tau^+) \to \pi^- (\pi^+) \pi^0 \nu_\tau (\bar{\nu}_\tau)$ and looked into the histogram of acoplanarity angle `$\phi$', which is the angle between the planes spanned by the $\rho^\pm$ decay (details of the observable construction is given in~\ref{app:obs}).  The choice of this particular decay channel is motivated to avoid the ambiguities arising from $\tau$ vertex reconstruction. 
To reproduce the plot, we have considered the Collin-Soper frame as described in subsection~\ref{subsec:anomalous_numerics}.  Our fig.~\ref{fig:anomalous_moment} nearly agrees with fig.~2 of Ref.~\cite{Banerjee:2022sgf}, and that is why it provides validation of our application and correctness of the polarimetric vectors as stored in \texttt{HepMC3} file. In the presence of hard photons, we consider the Mustraal frame for better phase space consideration and the plots appeared to be similar.

Of course, our method does not need to be limited to dipole moments; other new physics extra interactions can be monitored this way using $\tau$ spin. Advantage of the method of the new \texttt{C++} (\texttt{HepMC3} based) application is that information is taken from stored events and not on flight in generation, which is why for one sample, many variants of new physics modification can be introduced without the need for costly detector responses simulation.

\section{Summary} 
\label{sec:summary}
In this work, we have revisited and extended earlier applications of spin information in $\tau$-lepton physics, presenting them in a new and more user-friendly framework based on the \texttt{HepMC3} output of the \texttt{C++} version of \texttt{KKMCee}. This modernized format is directly applicable in experimental analyses, providing a practical bridge between precision Monte Carlo simulations and physics measurements. While the principles underlying our approach follow earlier works~\cite{ALEPH:2001uca,Banerjee:2022sgf}, the present implementation eliminates the need to access internal generator variables or methods, thereby lowering the technical barrier for users. The algorithms are designed to operate directly on \texttt{HepMC3}-stored events, ensuring portability and ease of integration into contemporary analysis frameworks.

To illustrate the usefulness of this approach, we have examined two representative examples: the extraction of $Z\tau\tau$ couplings from $\tau$ polarization measurements, and the study of anomalous $\tau$ dipole moments. In both cases, our results are consistent with previous literature, validating the correctness of the stored spin information and confirming the reliability of the new method. Further technical aspects of the implementation are provided in~\ref{app:tech}.

The primary objectives of this work are therefore twofold: first, to establish the correctness and completeness of the spin information recorded in \texttt{HepMC3} by \texttt{KKMCee}; and second, to present a flexible reweighting algorithm that enables the inclusion of additional short-distance interactions with characteristic scales comparable to or above the collider CM energy. This capability allows users to explore extensions of the Standard Model without the need to modify the generator itself, thereby broadening the scope of \texttt{KKMCee} for new-physics studies.

Looking forward, we intend to expand this framework toward a wider class of new-physics applications, including scenarios such as lepton-flavour violation~\cite{Hayasaka:2010np} and other non-standard $\tau$ couplings(such as~\cite{Ma:2001md,  Ganguly:2022qxs}). These directions are of particular interest for future collider programs, where high-statistics datasets and precise polarization measurements will play a central role. We hope that the results presented here will serve both as a validation of the current \texttt{KKMCee} implementation and as a foundation for experimental collaborations to exploit $\tau$ spin information in searches for physics beyond and of the Standard Model. Updated informations of our work and in particular application code explained in~\ref{app:stand_alone} can be found in~\href{https://th.ifj.edu.pl/kkmc-demos/index.html}{https://th.ifj.edu.pl/kkmc-demos/index.html}. You can find detailed information  on \texttt{KKMCee} on the main site, \href{https://kkmcee.docs.cern.ch/}{https://kkmcee.docs.cern.ch/}.
\newline

\centerline{\textbf{Acknowledgment}} 
This project was supported in part from funds of the National Science Centre, Poland, grant no. 2023/50/A/ST2/00224 and of COPIN-IN2P3 collaboration with LAPP-Annecy. We acknowledge Alexandr Yu. Korchin for the useful discussion during the project.  We are also thankful to Jacek Holeczek for his insights during the installation of the \texttt{KKMCee} program. 

\appendix 
\section{Technical details}
\label{app:tech}
\subsection{Attributes saved to \texttt{HepMC3}}
\label{app:attributes_saved_hepmc3}
The \texttt{HepMC3} output of \texttt{KKMCee} is extended with spin-aware attributes at event level and at particle level. This enables fast validation of the spin modeling and the helicity-only approximation directly in user analyses, without re-running the generator. The \texttt{HepMC3} ASCII output is enabled by setting \texttt{KKMCee} input parameter 100 to 1. The attributes are accessible via \texttt{HepMC3::ReaderAscii}, which parses each record into a \texttt{HepMC3::GenEvent} object. The newly added attributes are:

\begin{itemize}
\item \texttt{SpinWT} and \texttt{SpinWThelApprox} (event attributes): the full spin weight computed from the complete spin correlation matrix and its helicity approximated weight obtained after projecting onto the longitudinal (helicity) subspace. These entries allow physics users to test whether helicity separation is adequate for their analysis. The helicity approximated weight is computed as given in Eq.~\ref{eq:appx_spin_wt},

\begin{align}
    wt_{\text{appx}} &= \frac{1}{4}\sum_{m= \pm} \sum_{n=\pm} (h^+ \cdot s^m) (s^m \cdot R \cdot s^n) (h^- \cdot s^n) \\
    &= \frac{1}{4}\sum_{m,n=\pm}R_{mn}(1+mh_{\text{z}}^+)(1+nh_{\text{z}}^-)
\end{align}
where $R_{mn} = s^m \cdot R \cdot s^n$, since at $\tau$-rest frame $h_{\text{t}}^\pm = 1$. 

\item \texttt{approximateHelicity} (per-$\tau$ attributes) : For each event, the helicity tags $m,n \in \{+1, -1\}$ to $\tau^-$ and $\tau^+$ by sampling from the helicity posterior
\begin{equation}
    p(m,n|event) \propto R_{mn}(1+mh_{\text{z}}^+)(1+nh_{\text{z}}^-)
\end{equation}
normalized over (m,n). The stored per particle attribute \texttt{approximateHelicity} is thus a probabilistic tag, not a ground-truth helicity, which can be defined only for massless fermions.

\item \texttt{polarimetricInLabFrame} (per-$\tau$ attributes): For each $\tau^\pm$, the corresponding polarimetric four-vector $h^\pm$. They are defined in the laboratory frame. These vectors encode the full decay-channel analysing power and permit offline reweighting to alternative production dynamics without modifying the decay simulation.
\end{itemize}

\subsection{Standalone code for spin weight calculation}
\label{app:stand_alone}
The \texttt{F77} version of \texttt{KKMCee} already includes spin-weight calculations, with and without anomalous electric and magnetic dipole moment contributions for the $\tau$ lepton. However, that implementation is not standalone, so obtaining new weights requires rerunning the entire simulation. We provide a standalone \texttt{C++} implementation that computes the weights directly. This is only possible due to the inclusion of the new attributes in the \texttt{HepMC3} output file. A lightweight reader processes the \texttt{HepMC3} event record - including the newly added attributes and particle information and computes, event by event, the spin weights with and without the new couplings. The procedure mirrors the logic of  Ref.~\cite{Banerjee:2022sgf} and uses the same spin–correlation matrix $R_{ij}$ (there denoted $R_{ij}(s,\cos\theta)$). For completeness we summarize the algorithmic steps.

\begin{enumerate}
  \item Event attributes in the \texttt{HepMC3} record are read with the help   of \texttt{HepMC3::ReaderAscii}, provided with the                \texttt{HepMC3} package. Each event is represented as a                     \texttt{HepMC3::GenEvent}, which allows straightforward inspection and      (if needed) modification of the record. 

  \item The required information from \texttt{HepMC3::GenEvent} is retrieved using a custom \texttt{HepMCReader}, which can be readily adapted to specific decay topologies. The extracted elements are stored in the \texttt{EventInitilizers} class.
  
  \item The resulting \texttt{EventInitilizers} is passed to \texttt{AnomWt} to compute the spin weight; new-physics effects are enabled via 
  \texttt{AnomWt::setCouplings(Ar0, Ai0, Br0, Bi0)}, where \texttt{Ar0}, \texttt{Ai0}, \texttt{Br0}, and \texttt{Bi0} are the real and imaginary parts of the NP form factors $a(s)_{\mathrm{NP}}$ and $b(s)_{\mathrm{NP}}$ (i.e., \texttt{Ar0} $=\Re     (a)$, \texttt{Ai0} $=\Im (a)$, \texttt{Br0} $=\Re (b)$, \texttt{Bi0}        $=\Im (b)$). The \texttt{AnomWt::compute()} calculate both the $wt_{SM}$ (wtSPIN0) and $wt_{NP}$ (wtSPIN) in the same call. Outputs are available through the corresponding getter functions. In the SM case, the form factors are kept zero.
  
  \item Reconstruct the $\tau$ pair 4–momenta $p_+$ and $p_-$ from the event
    record and form $Q=p_++p_-$. Boost all relevant 4–vectors (the two
    polarimetric vectors $h_\pm$ and the beam directions.) to the $\tau^+\tau^-$
    rest frame defined by $Q$.
    
  \item Fix axes. In the pair rest frame we align the $\tau$ flight directions
     with $\pm\hat{z}$ by two rigid rotations: (i) an azimuthal rotation that brings the chosen $\tau$ into the $x$–$z$ plane, (ii) a polar rotation that points it along $+z$ while the partner points to $-z$. 
     
  \item To minimise ISR–induced biases, in the $\tau^+ \tau^-$ rest frame we support two standard production axes. 
  \begin{itemize}
      \item \textbf{Collins-Soper}: set \texttt{frameOption=1} to select this production axis.

      \item \textbf{Mustraal}: evaluate with \texttt{frameOption=2} and \texttt{frameOption=3}, then take a probability-weighted average using Eq.~\ref{eq:probs}.
  \end{itemize}

  The choice is user-selectable and propagated consistently through the spin-weight calculation.

  \item Boost the polarimetric vectors ($h_+$ and $h_-$) to the respective $\tau$ rest frame giving $h_t =1$.

  \item The spin correlation coefficients $R_{ij}$ is obtained following Ref.~\cite{Banerjee:2022sgf} for both the standard model case ($R0$) and with anomalous dipole couplings ($R$). 

  \item Using the time-component as the spin-averaged normalization, the production (matrix element) reweight is 
  \begin{equation}
  wt_{\text{ME}} = \frac{R_{tt}}{R0_{tt}}    
  \end{equation}
  
  \item For every event in the \texttt{HepMC3} record, we compute the following weights:
\begin{equation}
    wt_{\text{spin}}^{0} =  \sum_{i,j} \frac{R0_{ij}}{R0_{tt}}h_+^{i}h_-^{j}     
\end{equation}  
\begin{align}
    wt_{\text{spin}} &= \sum_{i,j} \frac{R_{ij}}{R_{tt}}h_+^ih_-^j  \nonumber \\
    wt &= wt_{\text{ME}} \times wt_{\text{spin}} / wt^0_{\text{spin}}
    \label{eqn:wtspin}
\end{align}

where conventions of frame orientation and $\tau$ handeness have to be adjusted. we effectively multiply $h^i_\pm$
components by (1,-1,1,-1), that is rotate by angle $\pi$ around second axis. In Eq.~\ref{eqn:wtspin} we divide by the Standard Model weight $wt_{\text{spin}}^0$ because the events were generated with the SM matrix element already applied. The ratio is therefore the pure reweighting factor that moves an event from the SM hypothesis to the new-physics one, isolating the NP effect and avoiding double counting of the SM contribution. The application format can be obtained from the website~\href{https://th.ifj.edu.pl/kkmc-demos/index.html}{https://th.ifj.edu.pl/kkmc-demos/index.html}.
\end{enumerate}

\label{sec:techn}

\subsection{Observable Construction}
\label{app:obs}

\begin{enumerate}
      \item The acoplanarity angle $\phi$ is defined in the $\rho^+\rho^-$     rest frame as the oriented angle between the decay planes spanned by $(\pi^+,\pi^0)$ and $(\pi^-,\pi^0)$. Let the unit normals be,
\[
\hat{n}_+ = \frac{\vec{p}_{\pi^+}\times  \vec{p}_{\pi^0}}{|\vec{p}_{\pi^+}\times \vec{p}_{\pi^0}|},
\qquad
\hat{n}_- = \frac{\vec{p}_{\pi^-}\times \vec{p}_{\pi^0}}{|\vec{p}_{\pi^-}\times \vec{p}_{\pi^0}|},
\]
and choose as reference direction the $\rho^-$ flight direction,
\[
\hat{r} = \frac{\vec{p}_{\rho^-}}{|\vec{p}_{\rho^-}|}
= \frac{\vec{p}_{\pi^-}+\vec{p}_{\pi^0}}{| \vec{p}_{\pi^-}+\vec{p}_{\pi^0}|}.
\]
Then
\[
\phi=\cos^{-1}\!\big(\hat{ n}_+\!\cdot\!\hat{n}_-\big)\in[0,\pi].
\]
and we map to $[0,2\pi]$ by setting $2\pi- \phi \to \phi$ if $\hat{r} \cdot (\hat{n_+}\times \hat{n_-})$ is negative.
\end{enumerate}

\bibliographystyle{unsrt} 
\bibliography{reference}

\begin{thebibliography}{10}

\bibitem{ALEPH:2001uca}
A.~Heister et~al.
\newblock {Measurement of the tau polarization at LEP}.
\newblock {\em Eur. Phys. J. C}, 20:401--430, 2001.

\bibitem{Eberhard:1989ve}
P.~H. Eberhard, B.~van Eijk, J.~Fuster, S.~Jadach, A.~M. Lutz, E.~Richter-Was,
  P.~Rosselet, O.~Schneider, and Z.~Was.
\newblock {THE tau POLARIZATION MEASUREMENT AT LEP}.
\newblock In {\em {LEP Physics Workshop}}, 10 1989.

\bibitem{L3:1998oan}
M.~Acciarri et~al.
\newblock {Measurement of tau polarization at LEP}.
\newblock {\em Phys. Lett. B}, 429:387--398, 1998.

\bibitem{DELPHI:1999yne}
P.~Abreu et~al.
\newblock {A Precise measurement of the tau polarization at LEP-1}.
\newblock {\em Eur. Phys. J. C}, 14:585--611, 2000.

\bibitem{CMS:2023mgq}
Aram Hayrapetyan et~al.
\newblock {Measurement of the {\ensuremath{\tau}} lepton polarization in Z
  boson decays in proton-proton collisions at $ \sqrt{s} $ = 13 TeV}.
\newblock {\em JHEP}, 01:101, 2024.

\bibitem{Banerjee:2022sgf}
Sw. Banerjee, A.~Yu. Korchin, and Z.~Was.
\newblock {Spin correlations in {\ensuremath{\tau}}-lepton pair production due
  to anomalous magnetic and electric dipole moments}.
\newblock {\em Phys. Rev. D}, 106(11):113010, 2022.

\bibitem{Richter-Was:2016mal}
E.~Richter-Was and Z.~Was.
\newblock {Separating electroweak and strong interactions in
  Drell{\textendash}Yan processes at LHC: leptons angular distributions and
  reference frames}.
\newblock {\em Eur. Phys. J. C}, 76(8):473, 2016.

\bibitem{Jadach:1998jb}
S.~Jadach, B.~F.~L. Ward, and Z.~Was.
\newblock {Coherent exclusive exponentiation CEEX: The Case of the resonant e+
  e- collision}.
\newblock {\em Phys. Lett. B}, 449:97--108, 1999.

\bibitem{Jadach:1999vf}
S.~Jadach, B.~F.~L. Ward, and Z.~Was.
\newblock {The Precision Monte Carlo event generator K K for two fermion final
  states in e+ e- collisions}.
\newblock {\em Comput. Phys. Commun.}, 130:260--325, 2000.

\bibitem{Davidson:2010rw}
N.~Davidson, G.~Nanava, T.~Przedzinski, E.~Richter-Was, and Z.~Was.
\newblock {Universal Interface of TAUOLA Technical and Physics Documentation}.
\newblock {\em Comput. Phys. Commun.}, 183:821--843, 2012.

\bibitem{Jadach:1998wp}
S.~Jadach, B.~F.~L. Ward, and Z.~Was.
\newblock {Global positioning of spin GPS scheme for half spin massive
  spinors}.
\newblock {\em Eur. Phys. J. C}, 22:423--430, 2001.

\bibitem{Collins:1977iv}
John~C. Collins and Davison~E. Soper.
\newblock {Angular Distribution of Dileptons in High-Energy Hadron Collisions}.
\newblock {\em Phys. Rev. D}, 16:2219, 1977.

\bibitem{Barberio:1993qi}
Elisabetta Barberio and Zbigniew Was.
\newblock {PHOTOS: A Universal Monte Carlo for QED radiative corrections.
  Version 2.0}.
\newblock {\em Comput. Phys. Commun.}, 79:291--308, 1994.

\bibitem{Hayasaka:2010np}
K.~Hayasaka et~al.
\newblock {Search for Lepton Flavor Violating Tau Decays into Three Leptons
  with 719 Million Produced Tau+Tau- Pairs}.
\newblock {\em Phys. Lett. B}, 687:139--143, 2010.

\bibitem{Ma:2001md}
Ernest Ma, D.~P. Roy, and Sourov Roy.
\newblock {Gauged L(mu) - L(tau) with large muon anomalous magnetic moment and
  the bimaximal mixing of neutrinos}.
\newblock {\em Phys. Lett. B}, 525:101--106, 2002.

\bibitem{Ganguly:2022qxs}
Sougata Ganguly, Sourov Roy, and Ananya Tapadar.
\newblock {Secluded dark sector and muon (g-2) in the light of fast expanding
  Universe}.
\newblock {\em JCAP}, 02:044, 2023.

\end{thebibliography}

\end{document}